\documentclass[prl,showpacs,amsmath,amssymb, twocolumn]{revtex4}
\usepackage{graphicx}
\usepackage{amssymb}
\usepackage{dcolumn}
\usepackage{bm}

\begin{document}
\title{Competing Ordered Phases in URu$_2$Si$_2$: Hydrostatic Pressure and Re-substitution}
\author{J. R. Jeffries}
\author{N. P. Butch}
\author{B. T. Yukich}
\author{M. B. Maple}
\affiliation{Department of Physics and Institute for Pure and
Applied
Physical Sciences,\\
University of California, San Diego, La Jolla, CA 92093}

\date\today

\begin{abstract}
A persistent kink in the pressure dependence of the ``hidden
order'' (HO) transition temperature of URu$_{2-x}$Re$_x$Si$_2$ is
observed at a critical pressure $P_c$=15 kbar for $0 \leq x \leq
0.08$. In URu$_2$Si$_2$, the kink at $P_c$ is accompanied by the
destruction of superconductivity; a change in the magnitude of a
spin excitation gap, determined from electrical resistivity
measurements; and a complete gapping of a portion of the Fermi
surface (FS), inferred from a change in scattering and the
competition between the HO state and superconductivity for FS
fraction.
\end{abstract}



\maketitle

Since its discovery over 20 years ago \cite{Palstra1985,
Schlabitz1986, Maple1986}, the moderately heavy fermion compound
URu$_2$Si$_2$ has been the focus of many theoretical and
experimental efforts designed to determine the elusive, hidden
order parameter associated with the phase transition occurring at
$T_0 \approx$ 17.5 K. The transition into this ``hidden order''
(HO) state is characterized by large anomalies (typical of
magnetic ordering) in specific heat, electrical resistivity,
thermal conductivity, and magnetization measurements
\cite{Palstra1985, Schlabitz1986, Maple1986, McElfresh1987,
Behnia2005, Sharma2006, Pfleiderer2006}; however, only a small
antiferromagnetic moment, insufficient to adequately explain the
entropy released during the transition, was detected in
low-temperature neutron diffraction experiments
\cite{Broholm1987}. In addition to the puzzling order parameter of
the HO state, URu$_2$Si$_2$ undergoes a transition into an
unconventional superconducting (SC) state, which coexists with
weak antiferromagnetism (AFM), at $T_c \approx$ 1.5 K. The
potential interplay between the two ordered phases of
URu$_2$Si$_2$ as well as the nature of the HO state are underlying
problems to our fundamental understanding of the properties of
this compound.

In an effort to explain the observed properties of URu$_2$Si$_2$,
several microscopic models have been proposed \cite{Santini1994,
Kiss2005, Okuno1998, Varma2006, Chandra2002, Ikeda1998,
Mineev2005}. In addition to the theoretical pursuits, many varied
experimental techniques have been employed to confirm and/or
constrain the proposed models; however, the experimental results
fail to converge upon an encompassing microscopic description of
the ordered states of URu$_2$Si$_2$, but do provide valuable
insight when contextually analyzed. Low-temperature neutron
diffraction measurements as a function of magnetic field provide
evidence that the order parameter of the HO state must break
time-reversal symmetry \cite{Bourdarot2003}, and recent inelastic
neutron scattering measurements reveal gapped spin excitations at
incommensurate wavevectors \cite{Wiebe2007}. Thermal transport
measurements are consistent with the opening of a gap at the Fermi
surface (FS), as previously suggested by optical conductivity and
specific heat studies \cite{Bonn1988, Maple1986}, depleting
carriers and reducing electron-phonon scattering \cite{Behnia2005,
Sharma2006}. These exemplary measurements tend to converge upon a
description of the HO state invoking the presence of a FS
instability such as a spin density wave (SDW), further suggested
by high-field measurements intimating the itinerant nature of the
HO state \cite{Kim2003}.

The application of pressure to URu$_2$Si$_2$ further convolutes
the discussion of the nature of the hidden order parameter as well
as the persistence of the SC state. While the HO transition
temperature $T_0$ was seen to increase with applied pressure
\cite{McElfresh1987}, the SC state was found to be suppressed;
however, the reported critical pressures for superconductivity
vary greatly from 4-15 kbar, possibly due to disparities in sample
quality or high-pressure conditions between experiments
\cite{Sato2006, Amitsuka2007, McElfresh1987, Knebel2007}.
Pressure-dependent neutron scattering, NMR, and $\mu$SR
measurements all indicate an abrupt increase in the size of the
ordered moment, with the magnitude of the moment saturating to a
value consistent with bulk AFM above a currently disputed critical
pressure \cite{Amitsuka1999, Bourdarot2005, Matsuda2003,
Amitsuka2003}. The interpretation of this increase in moment is
currently unresolved, but can be divided into two prevailing
descriptions: an inhomogeneous, phase separated scenario where the
increase in the moment is due to an increase in volume fraction of
the observed, high-pressure moment \cite{Matsuda2003,
Amitsuka2003}; and a scenario involving two order parameters
describing the staggered magnetization and the HO state, with the
details governed by the coupling between order parameters
\cite{Bourdarot2005, Mineev2005}.

With the progression of experimental and theoretical
investigations into the ordered states of URu$_2$Si$_2$, now is a
critical time to clarify and categorize experimental observations
in order to enhance the understanding of the underlying phenomena.
The URu$_{2-x}$Re$_x$Si$_2$ system provides a unique opportunity
to study the effects of pressure on a HO state whose ambient
pressure transition temperature is reduced with rhenium
concentration $x$ \cite{Bauer2005}. In this letter we report new,
comprehensive high-pressure electrical resistivity measurements on
single crystal samples of the URu$_{2-x}$Re$_x$Si$_2$ system,
emphasizing with greater clarity than previous studies the
pressure dependence of the HO state and its relation to the field
and pressure dependence of the SC state. A change is observed in
the magnon dispersion gap along with a systematic evolution of the
scattering processes in URu$_2$Si$_2$. In addition, using a
framework developed by Bilbro and McMillan \cite{Bilbro1976}, the
competition for FS fraction between the HO and SC states is
quantified, engendering a consistent depiction of the pressure
dependence of the ordered states of URu$_2$Si$_2$.

\begin{figure}
\begin{center}\leavevmode
\includegraphics[scale=0.95]{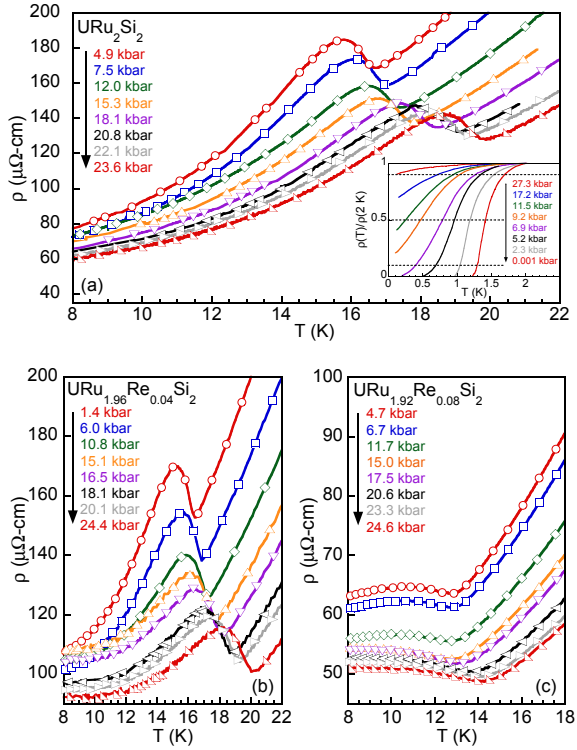}
\caption{(color online) (a) $\rho(T, P)$ of URu$_2$Si$_2$ as a
function of $T$. Inset: normalized electrical resistivity,
$\rho(T)/\rho$(2 K), vs. $T$ near $T_c$ (dashed lines defined in
the text). $\rho(T, P)$ data for specimens of
URu$_{2-x}$Re$_x$Si$_2$ with $x$=0.04 (b) and 0.08
(c).}\label{CombinedHOSC}
\end{center}
\end{figure}

Single crystals of URu$_{2-x}$Re$_x$Si$_2$ with $x$=0, 0.01, 0.02,
0.04, 0.06, and 0.08 were grown using the Czochralski method and
then annealed at 900 $^{\circ}$C for 7 days.  The Laue method was
used to orient the single crystals, which were subsequently spark
cut and polished into electrical resistivity specimens. Electrical
resistivity measurements under pressure were performed with a
beryllium-copper, piston-cylinder cell using a Teflon capsule
filled with a 1:1 mixture of n-pentane:isoamyl alcohol as the
pressure-transmitting medium to ensure hydrostatic conditions
during pressurization at room temperature.  The pressure in the
sample chamber was calibrated from the inductively determined,
pressure-dependent superconducting critical temperature of a lead
manometer.

\begin{figure}
\begin{center}\leavevmode
\includegraphics[scale=0.95]{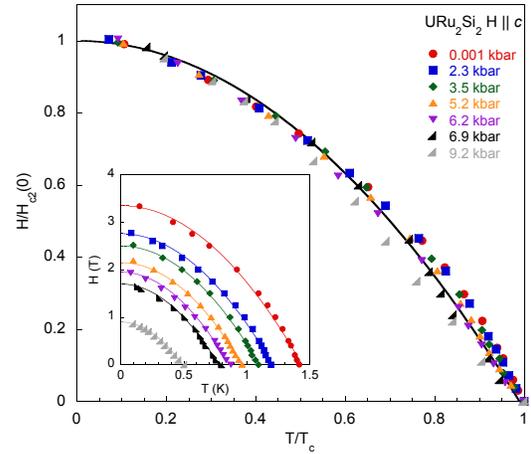}
\caption{(color online) Scaled critical field curves,
$H_{c2}(T)/H_{c2}(0)$ vs. $T/T_c$, of the SC state of
URu$_2$Si$_2$ for $P \leq$ 9.2 kbar and $H\parallel$ \textit{c}.
The solid line is a guide to the eye. Inset: $H_{c2}(T)$ vs $T$ of
the SC state. The solid lines are fits to a parabolic expression
yielding values of $H_{c2}(0)$.}\label{CriticalFields}
\end{center}
\end{figure}

Displayed in Figure \ref{CombinedHOSC}(a) are representative
electrical resistivity $\rho(T)$ data illustrating the evolution
of $T_0$, defined as the local minima occurring between
approximately 16.5 K and 20 K, with applied pressure. The data
show no qualitative difference in appearance as the characteristic
trough and peak structure of the HO transition is preserved to
high pressures. The inset of Figure \ref{CombinedHOSC}(a) shows
the electrical resistivity normalized at 2 K, $\rho(T)/\rho$(2 K),
emphasizing the pressure dependence of the SC transition. The
horizontal dashed lines represent--from top to bottom--90\%, 50\%,
and 10\% of the normal state value, with the SC critical
temperature $T_c$ defined as the temperature at which
$\rho(T)/\rho$(2 K)$=$0.5. Complete SC transitions were seen up to
nearly 7 kbar, and transitions to the 50\% value were observed up
to approximately 13 kbar; SC fluctuations persisted to the highest
pressures measured, although the roles of sample or pressure
inhomogeneities are undetermined. Figures \ref{CombinedHOSC}(b)
and (c) display $\rho(T)$ data near $T_0$ for specimens of
URu$_{1.96}$Re$_{0.04}$Si$_2$ and URu$_{1.92}$Re$_{0.08}$Si$_2$,
respectively.  While the absolute value of $T_0$ and the relative
value of the height of the transition in $\rho$ are altered with
$x$, the qualitative shape of the feature at $T_0$ is unchanged,
suggesting that, while Re-substitution suppresses $T_0$, the
physical mechanism responsible for the HO state persists. The
$x$-dependent changes in the magnitude of the resistivity possibly
occur from a change in the relative contribution of impurity
scattering with respect to the scattering intrinsic to the HO
state.

In addition to the pressure dependence of the SC state seen in the
inset of Figure \ref{CombinedHOSC}(a), field-dependent
measurements of the SC state were performed for $P<9.2$ kbar with
$H\parallel$\textit{c} (inset of Figure \ref{CriticalFields}). The
determined critical field curves were fit by a semi-empirical,
parabolic expression to extract $H_{c2}(0)$, the zero-temperature
upper critical field. Using these values of $H_{c2}(0)$, the
scaled critical field curves $H_{c2}(T)/H_{c2}(0)$ were plotted as
a function of reduced temperature $T/T_c$ (Figure
\ref{CriticalFields}). The data scale very close to one another
using these simple criteria, seemingly indicating that the
mechanisms governing the field dependence of the SC state remain
unchanged up to the critical pressure.

\begin{figure}
\begin{center}\leavevmode
\includegraphics[scale=0.95]{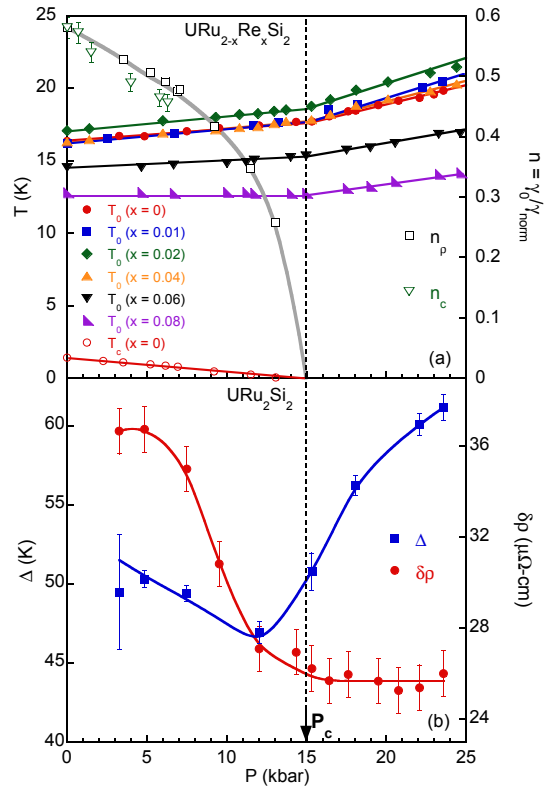}
\caption{(color online) (a) T-P phase diagram (left axis) of
URu$_{2-x}$Re$_x$Si$_2$ showing the evolution of the HO and SC
states. An $x$-independent kink in $T_0(P)$ is visible at $P_c$ =
15 kbar. The fraction of the FS left ungapped by the HO
transition, $n \equiv \gamma_0/\gamma_{norm}$ (right axis), vs.
$P$ as determined from $T_0$ and $T_c$, $n_{\rho}$ (open black
squares), and as estimated from previous $C(T,P)$ measurements
\cite{Fisher1990}, $n_c$ (open, inverted green triangles). (b) Gap
in the magnon dispersion $\Delta$ (left axis) from fits to
$\rho(T)$ below $T_0$. $\delta\rho$ (right axis) as defined in the
text. The lines are guides to the eye. The vertical dashed line
marks the critical pressure $P_c$ = 15 kbar.}\label{CombinedPhase}
\end{center}
\end{figure}

For all $x$ measured, $T_0$ and $T_c$ are plotted as a function of
$P$ in Figure \ref{CombinedPhase}(a), where $T_0$ evinces a
distinct kink in its pressure dependence at 15 kbar, regardless of
the ambient pressure value of $T_0$.  The presence of this kink,
when taken within the context of the analysis of Mineev and
Zhitomirsky \cite{Mineev2005}, indicates a scenario where there
exists no coupling between the order parameters of the HO state
and that of the high-pressure AFM phase, and furthermore suggests
that sample impurities or uncharacterized strains may be culpable
for the observed moment at low pressures. The persistence of this
kink and its static position upon reducing $T_0$ with increased
$x$ is consistent with a vertical or nearly vertical HO/AFM
transition occurring at $P_c$=15 kbar; as this transition is
indirectly probed, there can be no unambiguous determination as to
the degree of its order. This purported HO/AFM transition has been
directly observed in other measurements \cite{Motoyama2003,
Bourdarot2005, Amitsuka2007}; however, the reported critical
pressures are much lower than 15 kbar, typically near 7 kbar. This
discrepancy in $P_c$ could be due to sample dependence or
potential non-hydrostatic conditions present in previous
experiments utilizing a mixture of Fluorinert FC70/77, which
remains liquid (hydrostatic) only up to approximately 8 kbar
\cite{Sidorov2005}. In addition to the persistent kink at 15 kbar,
$T_c$ for URu$_2$Si$_2$ is suppressed very near $P_c$, consistent
with previous results suggesting that superconductivity and AFM
are mutually exclusive \cite{Knebel2007}. In fact, if
non-hydrostatic conditions are responsible for the early onset of
AFM at $P<$15 kbar, then one would expect the SC state to be
suppressed at these lower pressures as reported by several
researchers \cite{Sato2006, Amitsuka2007}.

It was suggested previously that the HO transition partially gaps
a portion of the FS with the remaining ungapped portion undergoing
superconductivity at low temperatures \cite{Maple1986}. In this
scenario, the existence of superconductivity is predicated upon
the incomplete FS gap induced by the onset of HO, and, as such,
the two ordered states compete for FS fraction. Bilbro and
McMillan proposed a model to quantitatively analyze the effects of
ordered phases competing for FS fraction \cite{Bilbro1976}:
\begin{eqnarray}
T_{c0}&=&T_c(P)^{n(P)} T_0(P)^{1-n(P)}\label{BilbroMcMillan1},
\end{eqnarray}
\noindent where $T_{c0}$ is the value of the SC critical
temperature in the absence of a high-temperature, FS-gapping
transition and $n(P)$ is a measure of the residual ungapped
FS--which can be determined from specific heat $C(T)$ measurements
as $n\equiv\gamma_0/\gamma_{norm}$, the ratio of the electronic
specific heat coefficient above $T_c$ to that above $T_0$.  Using
the previously determined value of $n(0)$=0.58 \cite{Maple1986},
and evaluating Equation \ref{BilbroMcMillan1} at ambient pressure
results in a value of $T_{c0}$=3.9 K, which can then be used along
with the values of $T_0(P)$ and $T_c(P)$ to quantify the fraction
of the FS that is gapped by the HO transition. The results of this
analysis (labelled $n_{\rho}$) are contained in Figure
\ref{CombinedPhase}(a)--where the HO transition would appear,
invoking a reasonable extrapolation, to completely gap its portion
of the FS near $P_c$ \cite{Footnote}, thus suppressing
superconductivity in the vicinity of $P_c$ as seen. Also plotted
in Figure \ref{CombinedPhase}(a) are extracted results for $n(P)$
from $C(T)$ data under pressure (labelled $n_c$)
\cite{Fisher1990}, which are in excellent agreement with the
analysis of the $\rho(T)$ data presented herein, although $C(T)$
data to higher pressures would be desirable to confirm this
postulate.

The electrical resistivity of Figure \ref{CombinedHOSC}(a) was fit
from approximately 2 K up to 90\% of $T_0$ with the previously
employed expression \cite{Palstra1986, Andersen1980,
McElfresh1987}:
\begin{eqnarray}
\rho(T)&=&\rho_{0}+AT^2+B\frac{T}{\Delta}\left(1+\frac{T}{\Delta}\right)e^{(-\Delta/T)},
\label{ReMagnonFit}
\end{eqnarray}
\noindent which accounts for the residual resistivity $\rho_{0}$,
a heavy Fermi liquid term, and scattering from gapped spin
excitations. The magnitude of the magnon gap $\Delta$ was found to
undergo a change in its pressure dependence near $P_c$, as shown
in Figure \ref{CombinedPhase}(b), where the error bars result from
the fitting algorithm used. Within a SDW formalism, this magnon
dispersion gap $\Delta$ is associated with the magnitude of the FS
gap \cite{Gruner1994}; and, although other conceivable
pressure-dependent parameters are involved in the relationship,
$\Delta(P)$ is an indirect probe of the pressure-dependent
evolution of the FS. Also shown in Figure \ref{CombinedPhase}(b)
is the magnitude of the HO anomaly $\delta\rho$, defined by
extrapolating the temperature dependences above and below the HO
transition to $T_0$ and evaluating the difference in the resultant
electrical resistivity; the ascribed error bars result from small
differences in permissible extrapolations. The quantity
$\delta\rho$ decreases with pressure up to $P_c$, after which it
remains constant.  This behavior can be understood within the
context of the fraction of FS left ungapped by the HO transition:
with applied pressure, the fraction of FS gapped by the HO
transition grows larger and reduces the number of available states
into which quasiparticles can scatter, resulting in a reduction in
the scattering and a consequent reduction in $\delta\rho$; at and
above $P_c$, the FS is completely gapped and the scattering
processes along with $\delta\rho$ become constant.

The salient features of the results presented herein can be
summarized as follows: (1) the HO state and its associated
qualitative features persist in URu$_{2-x}$Re$_x$Si$_2$,
suggesting that Re substitution does little to alter the HO state;
(2) a distinct kink in $T_0(P)$ is seen at $P_c$=15 kbar and the
kink is independent of the value of $T_0(0)$ as modified by
increasing $x$; (3) the HO state coincides with superconductivity,
and the mechanism for superconductivity appears unaltered
according to the variations of $T_c$ and $H_{c2}(T)$ with $P$; (4)
the HO and SC states compete for FS fraction, with the former
occupying 100\% of its portion of FS near 15 kbar; (5) the gap
$\Delta$ associated with magnetic excitations inferred from
$\rho(T,P)$ measurements evinces a change in behavior near 15
kbar. It would appear likely that at the critical pressure of
$P_c$=15 kbar, URu$_2$Si$_2$ undergoes a distinct HO/AFM
transition, although the degree of order and nature of the
transition remain uncertain. Furthermore, this HO/AFM transition
seemingly occurs along a vertical or near vertical phase boundary
near 15 kbar. The coincidence of the HO/AFM boundary, the fully
gapped portion of the FS where HO resides, and the change in
$\Delta$ is not currently understood. No particular microscopic
model has been invoked to discuss the results, although the
simultaneous presence of a spin and FS gap strongly favor the
formation of a SDW-like FS instability at $T_0$.

Crystal growth was sponsored by the U.S. Department of Energy
(DOE) under Research Grant \# DE-FG02-04ER46178. Measurements were
sponsored by the National Nuclear Security Administration under
the Stewardship Science Academic Alliances program through DOE
Research Grant \# DE-FG52-06NA26205.

\end{document}